Sequential Optimization Numbers and Conjecture about Edge-Symmetry and Weight-Symmetry Shortest Weight-Constrained Path

Zile Hui

East China Normal University, 51174500096@stu.ecnu.edu.cn

This paper defines multidimensional sequential optimization numbers and prove that the unsigned Stirling numbers of first kind are *1*-dimensional sequential optimization numbers. This paper gives a recurrence formula and an upper bound of multidimensional sequential optimization numbers. We proof that the *k*-dimensional sequential optimization numbers, denoted by $O_k(n,m)$, are almost in $\{O_k(n,\alpha)\}$, where $\alpha \in \left[1, \lceil ek \cdot \log(n-1) + \frac{e\pi^2}{6}(2^k - 1) \rceil + M_1 \right]$, $n$ is the size of *k*-dimensional sequential optimization numbers and $M_1$ is large positive integer. Many achievements of the Stirling numbers of first kind can be transformed into the properties of *k*-dimensional sequential optimization numbers by *k*-dimensional extension and we give some examples.

Shortest weight-constrained path is NP-complete problem [1]. In the case of edge symmetry and weight symmetry, we use the definition of the optimization set to design *2*-dimensional Bellman-Ford algorithm to solve it. According to the fact that $P_1(n, m > M) \leq e^{-M_1}$, where $M = \lceil e\log(n-1) + e \rceil + M_1$, $M_1$ is a positive integer and $P_1(n,m)$ is the probability of 1-dimensional sequential optimization numbers, this paper conjecture that the probability of solving edge-symmetry and weight-symmetry shortest weight-constrained path problem in polynomial time approaches 1 exponentially with the increase of constant term in algorithm complexity. The results of a large number of simulation experiments agree with this conjecture.

## 1 INTRODUCTION

Stirling numbers have been paid lots of interests since they were introduced in 1730's. There is a known recurrence formula for the unsigned Stirling numbers of first kind but there is no known closed-form expression so far [6,7]. This paper proof that the unsigned Stirling numbers of first kind are *1*-dimensional sequential optimization numbers. An upper bound of sequential optimization numbers is proved by using the recurrence formula of the sequential optimization numbers and it shows that the sequence is mostly concentrated in the smaller part. This paper also gives some properties of sequential optimization numbers.

In the short time since introduction of NP-complete in the early 1970's, this term has been recognized as one of the most important unsolved problem in mathematics and theoretical computer science. In many classical algorithms, one of the basic steps is the selection of optimal terms which are usually one dimension. By means of the definition of optimization set, steps for selecting optimization items can be extended from one dimension to multiple dimensions. On this basis, *2-dimensional* Bellman-Ford algorithm is designed to solve edge-symmetry and weight-symmetry shortest weight-constrained path problem. According to the properties of *1*-dimensional sequential optimization numbers, this paper conjecture that the probability of solving the problem in polynomial time is almost 1.

## 2 SEQUENTIAL OPTIMIZATION NUMBERS

We define the optimization set in Definition 2.1 and define the *k*-dimensional sequential optimization numbers in Definition 2.2. Then, we give an expression, which is not closed-form, a recurrence formula, some properties and applications of sequential optimization numbers. At the end of this chapter, we give an upper bound of sequential optimization numbers and some conclusions derived from this upper bound.

**Definition 2.1.** Let $\boldsymbol{x}(x_1, x_2, \ldots, x_k)$, $\boldsymbol{y}(y_1, y_2, \ldots, y_k)$ be $k$-dimensional vectors and $\boldsymbol{R}(R_1, R_2, \ldots, R_k)$ be $k$-dimensional relation vector. If for all $i \in \{1,2,\ldots,k\}$, $(x_i, y_i) \in R_i$, then $\boldsymbol{x}$ is said to be related to $\boldsymbol{y}$ by $\boldsymbol{R}$, denoted by $(\boldsymbol{x}, \boldsymbol{y}) \in \boldsymbol{R}$. Let $U$ be a set of vectors, $\boldsymbol{R}$ be a relation vector and $A \subseteq U$, if for $\forall \boldsymbol{u} \in U$, $\exists \boldsymbol{a} \in A$ imply $(\boldsymbol{a}, \boldsymbol{u}) \in \boldsymbol{R}$ or $\boldsymbol{a} = \boldsymbol{u}$, then $A$ is said to be a majorization set of $U$ by $\boldsymbol{R}$. If $B$ is a majorization set of $U$ by $\boldsymbol{R}$ and $|B|$ is a minimum of all the cardinalities of majorization sets, then $B$ is said to be an optimization set of $U$ by $\boldsymbol{R}$, denoted by $B \underset{\subseteq}{\overset{O\boldsymbol{R}}{}} U$, and $|B|$ is said to be the weight of $U$ by $\boldsymbol{R}$, denoted by $W_{o\boldsymbol{R}}(U)$.

We can find similar selection strategies of optimization set in many existing studies [2,5]. It can be easily proved that this strategy can include all possible optimal terms and exclude impossible optimal terms in the iteration of the algorithm presented later in this paper. However, this selection strategy can easily lead to exponential time complexity.

**Definition 2.2.** Let $\boldsymbol{a_1}(1, a_{11}, a_{12}, \ldots a_{1k})$, $\boldsymbol{a_2}(2, a_{21}, a_{22}, \ldots a_{2k})$, …, $\boldsymbol{a_n}(n, a_{n1}, a_{n2}, \ldots a_{nk})$ be $n$ $k+1$-dimensional vectors and for all $j \in \{1,2,\ldots,k\}$, $a_{1j}, a_{2j}, \ldots a_{nj}$ are $1,2,\ldots,n$, respectively. There are $n!^k$ ways. Let $\boldsymbol{R}(R_1, R_2, \ldots, R_k) = (<, <, \ldots, <)$ be $k$-dimensional relation vector, $U = \{\boldsymbol{a_1}, \boldsymbol{a_2}, \ldots, \boldsymbol{a_n}\}$, $\boldsymbol{b_{ij}} = (i, a_{ij})$, $U_j = \{\boldsymbol{b_{1j}}, \boldsymbol{b_{2j}}, \ldots, \boldsymbol{b_{nj}}\}$, $S_j$ is an optimization set of $U_j$ by $(<, R_j)$, where $i \in \{1,2,\ldots,n\}$ and $j \in \{1,2,\ldots,k\}$. Let $S \subseteq U$, for all $j \in \{1,2,\ldots,k\}$, if at least one $\boldsymbol{b_{ij}} \in S_j$, then $\boldsymbol{a_i} \in S$, otherwise, $\boldsymbol{a_i} \notin S$, where $i \in \{1,2,\ldots,n\}$. $S$ is said to be a sequential optimization set of $U$ by $\boldsymbol{R}$, denoted by $S \underset{\subseteq}{\overset{O\boldsymbol{R}}{}} U$. $|S|$ is said to be sequential optimization weight of $U$ by $\boldsymbol{R}$, denoted by $W_{O\boldsymbol{R}}(U)$. The numbers of ways that $W_{O\boldsymbol{R}}(U) = m$ are said to be $k$-dimensional sequential optimization numbers, denoted by $O_k(n, m)$. For all $n = 0$ and $k \geq 0$, we define that

$$O_k(n, m) = \begin{cases} 1, & m = 0 \\ 0, & m > 0 \end{cases}$$

For all $n > 0$ and $k = 0$ we define that

$$O_k(n, m) = \begin{cases} 1, & m = 1 \\ 0, & otherwise \end{cases}$$

**Theorem 2.1.** For all $k, n \in N$,

$$O_k(n, m) = \begin{cases} (n-1)!^k, & m = 1, n > 0 \\ (n-1)!^k \sum \prod_{i=1}^{m-1} \dfrac{j_i^k - (j_i - 1)^k}{(j_i - 1)^k}, & m \in \{2, 3, \ldots n\} \\ 1, & m = 0, n = 0 \\ 0, & otherwise \end{cases}$$

where $j_1, j_2, \cdots, j_{m-1}$ are all $m$-1-combinations of $\{2, 3, \ldots, n\}$.

**Theorem 2.2.** *For all $m, n \in N$ and $k \in N^+$, the recurrence formula of k-dimensional sequential optimization numbers is*

$$O_k(n+1, m+1) = n^k O_k(n, m+1) + [(n+1)^k - n^k] O_k(n, m)$$

*and the boundary condition is*

$$O_k(n, m) = \begin{cases} 0, & m = 0, n > 0 \text{ or } m > n \\ 1, & m = 0, n = 0 \end{cases}$$

We also notice that the recurrence formula works for all $m, n, k \in N$, if we define $0^0 = 0$.

We give some properties and applications of sequential optimization numbers below and we only prove part of them in this paper.

**Lemma 2.1.**

$$\sum_{m=1}^{n} O_k(n, m) = n!^k$$

**Lemma 2.2.** *For k=1, $O_1(n, m) = s_u(n, m)$, where $s_u(n, m)$ are the unsigned Stirling numbers of first kind.*

**Lemma 2.3.** *For all $n \in N^+$ and $k \in N$, we define that*

$$x_k^{n\uparrow} = x[(2^k - 1)x + 1][(3^k - 2^k)x + 2^k] \cdots \{[n^k - (n-1)^k]x + (n-1)^k\}$$

*and $O_k^u(n, m) = O_k(n, m)$. Then, we can get*

$$x_k^{n\uparrow} = \sum_{m=0}^{n} O_k^u(n, m) x^m$$

*and the zero points are $x = 0$ and $x = \frac{(m-1)^k - m^k}{(m-1)^k}$, where $m \in \{2, 3, \ldots, n\}$. We call $O_k^u(n, m)$ unsigned sequential optimization numbers.*

*For all $n \in N^+$ and $k \in N$, we define that*

$$x_k^{n\downarrow} = x[(2^k - 1)x - 1][(3^k - 2^k)x - 2^k] \cdots \{[n^k - (n-1)^k]x - (n-1)^k\}$$

*and $O_k^s(n, m) = (-1)^{n+m} O_k(n, m)$. Then, we can get*

$$x_k^{n\downarrow} = \sum_{m=0}^{n} O_k^s(n, m) x^m$$

*and the zero points are $x = 0$ and $x = \frac{m^k - (m-1)^k}{(m-1)^k}$, where $m \in \{2, 3, \ldots, n\}$. We call $O_k^s(n, m)$ signed sequential optimization numbers.*

**Instance 2.1.** *There are n boards and the sequence of them are fixed. Each board is painted one color and divided into k smaller boards. All the ith smaller boards of each board form a group and their height are $1, 2, \ldots, n$ respectively, where $i \in \{1, 2, \ldots, k\}$. Following the direction of the arrow, the number of ways that m colors can be seen is $O_k(n, m)$.*

We call it *k*-dimensional color boards problems. An example is shown in Figure 1.*a* and the number of colors that can be seen is shown in Figure 1.*b*. We find that *2*-dimensional color boards problems has been presented as sequence *A309053* in the On-Line Encyclopedia of integer Sequences (OEIS) and the author calculates this sequence by enumeration.

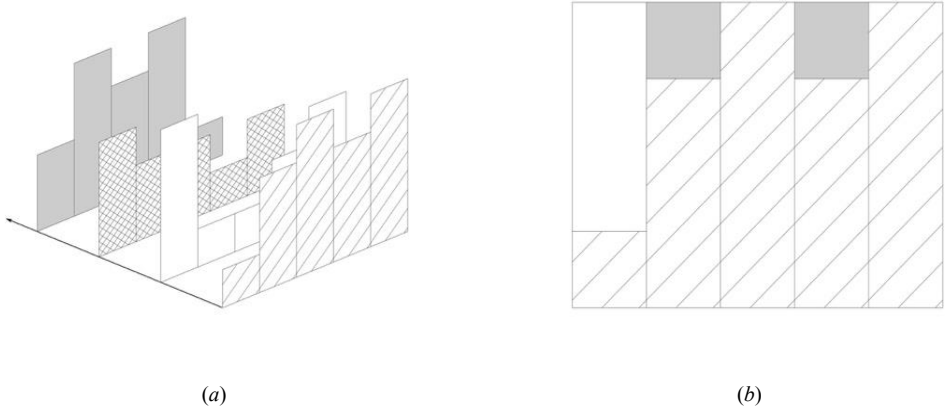

(a)　　　　　　　　　　　　　　　(b)

Figure 1: An example of *k*-dimensional color boards problems.

We can view Stirling numbers of first kind as a special case of *k*-dimensional sequential optimization numbers when *k*=1. Since Stirling numbers of first kind have been widely studied and a large number of achievements have been made, many of them can be transformed into the properties of *k*-dimensional sequential optimization numbers by *k*-dimensional extension.

**Theorem 2.3.** *For all $k \geq 1, n \geq 2$ and $m \geq 1$, we define*

$$O_{kmax}(n,m) = \frac{(n-1)!^k}{(m-1)!} \{\sum_{i=1}^{k}[C_k^i \cdot H_i(n-1)]\}^{m-1}$$

*where $H_i(n-1) = \sum_{j=1}^{n-1}\frac{1}{j^i}$. We can get $O_{kmax}(n,m) \geq O_k(n,m)$.*

**Theorem 2.4.** *Let $P_k(n,m) = \frac{O_k(n,m)}{n!^k}$ be the probability of $O_k(n,m)$ and $M = \lceil ek \cdot log(n-1) + \frac{e\pi^2}{6}(2^k-1) \rceil + M_1$, where $M_1$ is a positive integer. For all $n \geq 2$ and $k \geq 1$, We can get*

$$P_k(n, m > M) \leq e^{-M_1}$$

$P_k(n, m > M)$ decreases exponentially with the increase of $M_1$, so $O_k(n,m)$ are mostly concentrated in $\{O_k(n,\alpha)\}$, where $\alpha \in [1, M]$ and $M_1$ is a large positive integer.

**Lemma 2.4.** *In particular, for $k = 1, M = \lceil e\,log(n-1) + e \rceil + M_1$, $P_1(n, m > M) \leq e^{-M_1}$. $O_1(n,m)$, or $s_u(n,m)$, are mostly concentrated in $\{O_1(n,\alpha)\}$, where $\alpha \in [1, \lceil e\,log(n-1) + e \rceil + M_1]$. Let $M_2 = M_1 + 3$ and $O_1(n,m)$ are mostly concentrated in $\{O_1(n,\alpha)\}$, where $\alpha \in [1, \lceil e\,log(n-1) \rceil + M_2]$ and $M_2$ is a large positive integer.*

**Theorem 2.5.** *For all $k \geq 1, n \geq 2$,*

$$\frac{\sum_{i=1}^{n} O_{kmax}(n,i)}{\sum_{i=1}^{n} O_k(n,i)} \leq (\frac{n-1}{n})^k e^{\frac{\pi^2}{6}(2^k-1)}$$

*In particular, for all $k = 1, n \geq 2$,*

$$\frac{\sum_{i=1}^{n} O_{1max}(n,i)}{\sum_{i=1}^{n} O_1(n,i)} \leq \frac{(n-1)}{n} e^{H_1(n-1)-log\,(n-1)} \leq e$$

# 3 CONJECTURE ABOUT EDGE-SYMMETRY AND WEIGHT-SYMMETRY SHORTEST WEIGHT-CONSTRAINED PATH

We give the definition of *k*-dimensional optimization numbers and some properties of them. Then, we introduce the *2*-dimensional Bellman-Ford algorithm and use it to solve the edge-symmetry and weight-symmetry shortest weight-constrained path pronlem. We also analyze the complexity of the algorithm. At the end of this chapter, we present an analysis based on one hypothesis and propose the conjecture.

## 3.1 *K*-dimensional optimization numbers

**Definition 3.1.** *Let $\boldsymbol{c_1}(c_{11}, c_{12}, \dots c_{1k}), \boldsymbol{c_2}(c_{21}, c_{22}, \dots c_{2k}), \dots, \boldsymbol{c_n}(c_{n1}, c_{n2}, \dots c_{nk})$ be n k-dimensional vectors and $c_{1j}, c_{2j}, \dots, c_{nj}$ are $1, 2, \dots n$, respectively, where $j \in \{1, 2, \dots, k\}$. There are $n!^k$ ways. According to the definition of optimization sets, a certain way can get $n!$ same optimization sets by swapping $\boldsymbol{c_1} \sim \boldsymbol{c_n}$. So, for all $i \in \{1, 2, \dots, n\}$, we assume that $c_{i1} = i$, and discuss the remaining $n!^{k-1}$ ways. Let $\boldsymbol{R}(R_1, R_2, \dots, R_k) = (<, <, \dots, <)$ be k-dimensional relation vector and $U = \{\boldsymbol{c'_1}, \boldsymbol{c'_2}, \dots, \boldsymbol{c'_n}\}$, where $\boldsymbol{c'_1}(1, c_{12}, \dots c_{1k}), \boldsymbol{c'_2}(2, c_{22}, \dots c_{2k}), \dots, \boldsymbol{c'_n}(n, c_{n2}, \dots c_{nk})$. The numbers of ways that $W_{o\boldsymbol{R}}(U) = m$ are said to be k-dimensional optimization numbers, denoted by $o_k(n, m)$.*

We define this sequence in terms of optimization set. This sequence has been defined and its relation to the unsigned Stirling numbers of first kind has been explained [8]. Part of this sequence have been studied in some literature [3,4].

**Lemma 3.1.** *We can easily get $o_2(n, m) = O_1(n, m) = s_u(n, m)$.*

**Lemma 3.2.** *$\boldsymbol{R'} = (<, <, \dots, <)$ is k+1-dimensional relation vector and in the scenario of definition 2.2, for a certain way of $U$, $W_{o\boldsymbol{R'}}(U) = m_1$, $W_{o\boldsymbol{R}}(U) = m_2$, $C_1 \overset{o\boldsymbol{R'}}{\subseteq} U$, $C_2 \overset{O\boldsymbol{R}}{\subseteq} U$ and $\boldsymbol{c} \in U$. If $\boldsymbol{c} \notin C_1$, then $\boldsymbol{c} \notin C_2$, so that $m_1 \geq m_2$. Let $o_{k+1}(n, m \geq \gamma) = \sum_{i=\gamma}^{n} o_{k+1}(n, i)$ and $O_k(n, m \geq \gamma) = \sum_{i=\gamma}^{n} O_k(n, i)$, then $o_{k+1}(n, m \geq \gamma) \geq O_k(n, m \geq \gamma)$, where $\gamma \in \{0, 1, \dots, n\}$.*

**Lemma 3.3.** *$U = \{\boldsymbol{a_1}, \boldsymbol{a_2}, \dots \boldsymbol{a_n}\}$, $\boldsymbol{R} = (\leq, \leq)$, $\boldsymbol{a_i} = (x_i, y_i)$, $x_i$ are symmetric with each other and $y_i$ are symmetric with each other. We assume that $x_1 \leq x_2 \leq \dots \leq x_n$ and there are $n!$ different sorts of $y_i$. For $\boldsymbol{R'} = (<, <)$, $x_1 < x_2 < \dots < x_n$ and the relationship between $y_i$ is $<$, the number of ways that $W_{o\boldsymbol{R'}}(U) = m$ equals $o_2(n, m)$. For a certain sort, if we change $<$ to $\leq$ between $x_i$ and between $y_i$, then $W_{o\boldsymbol{R}}(U) \leq W_{o\boldsymbol{R'}}(U)$ is only possible. So, $W_{o\boldsymbol{R}}(U)$ are mostly concentrated in $\left[1, \lceil e \log(n-1) \rceil + M_2\right]$, where $M_2$ is a large positive integer.*

## 3.2 *2*-dimensional Bellman-Ford algorithm

We introduce the shortest weight-constrained path problem.

**Instance 3.1.** *Graph $G = (V, E)$, length $l(e) \in Z^+$, and weight $w(e) \in Z^+$ for each $e \in E$, specified vertices $s, t \in V$, positive integers $K, W$. Question is that is there a simple path in $G$ from $s$ to $t$ with total weight $W$ or less and total length $K$ or less?*

When $l(e)$ are symmetric with each other and $w(e)$ are symmetric with each other, We call it edge-symmetry and weight-symmetry shortest weight-constrained path problem.

We give the *2*-dimensional Bellman-Ford algorithm below. $OPT(i, v)$ denotes the smallest possible set of vectors composed of lengths and weights of *v-t* paths consisted of at most *i* edges and $M(i, v) = OPT(i, v)$ in the algorithm. Let $\boldsymbol{a_{vw}} = (l(e_{vw}), w(e_{vw}))$ and $OPT(i - 1, w) + \boldsymbol{a_{vw}}$ denotes the set of vectors that we get by adding $\boldsymbol{a_{vw}}$ and every element of $OPT(i - 1, w)$. $\boldsymbol{R} = (\leq, \leq)$. Let

$$OPT(i, v) \overset{o\boldsymbol{R}}{\subseteq} (OPT(i-1, v) \cup \bigcup_{w \in V} \{OPT(i-1, w) + \boldsymbol{a_{vw}}\}) \quad (1)$$

| ALGORITHM 1: *2*-dimensional Bellman-Ford algorithm |
|---|

Shortest-Path(G,s,t)
    n=number of nodes in G
    Multidimensional array M[0,1, … n-1,V]
    Dynamic array M[0,t]=(0,0) and for the other nodes v ∈ V, dynamic array M[0,v]=($\infty, \infty$)
    For i=1,2…,n-1
        For each v ∈ V in any order
          Compute M[i,v] using the formula (1)
        end
    end
Return M[n-1,s]

We give a way to update formula (1). $OPT(i - 1, v)$ is a optimization set. Take each element in $OPT(i - 1, w) + \boldsymbol{a_{vw}}$ in turn, determine whether it is related to every element in $OPT(i - 1, v)$ by **R** or the other way round and updata to form a new optimization set $OPT(i - 1, v)$. The last updata set is $OPT(i, v)$.

We assum that number of edges of every node is $O(n)$ and analyze the complexity of the algorithm. The complexity of step 0 is $O(n)$. After step 1 is executed, the size of dynamic array $M[1, v]$ is 1 and the complexity is $O(n^2)$. After step 2 is executed, the size of dynamic array $M[2, v]$ is $O(n)$ and the complexity is $O(n \sum_{i=1}^{n} i) = O(n^3)$ according to the updata method. And by analogy, after step n-2 is executed, the size of dynamic array $M[n - 2, v]$ is $O(n^{n-3})$ and the complexity is $O\left(n \sum_{i=n^{n-4}}^{n^{n-3}} i\right) = O(n^{2n-5})$. In the last step,only one point s needs to be updataed and the complexity is $O\left(\sum_{i=n^{n-3}}^{n^{n-2}} i\right) = O(n^{2n-4})$. So the complexity of the algorithm is $O(n^{2n-4})$, which is exponential. The algorithm is not supposed to be a good algorithm. Even with improvements, it is unlikely to be polynormial because that means solving NP-complete problem.

### 3.3 Hypothesis and conjecture

We Use *2*-dimensional Bellman-Ford algorithm to solve edge-symmetry and weight-symmetry shortest weight-constrained path problem. In simulation experiments, $l(e)$ are independent of each other and belong to the same probability distribution. $w(e)$ are independent of each other and belong to the same probability distribution. Let $L[i, v]$ be the size of $M[i, v]$ and $p_{ei} = \max(L[i, v])$, where $i \in \{0,1,,…,n - 1\}$ and $v \in V$. In a large number of simulation experiments, we find that the graph shown in Figure 2.*a* has larger $p_{ei}$ in the case of the same number of edges. Take the graph in Figure 2.*a* and Figure 2.*b* shows the relationship between $i$ and $p_{ei}$ in a simulation experiment. With the increase of $i$, $p_{ei}$ does not increase exponentially and is within an acceptable range. All other simulation experiment have similar results.

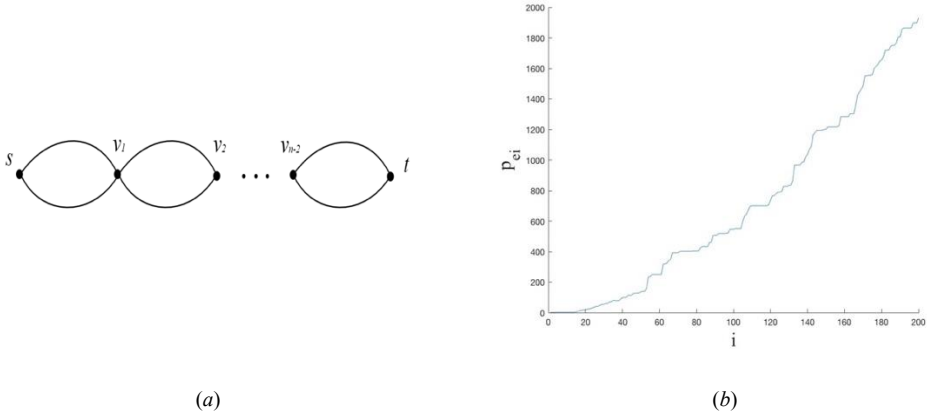

(a)  (b)

Figure 2: A simulation experiment of *2*-dimensional Bellman-Ford algorithm.

We present an analysis based on one hypothesis. When *2*-dimensional Bellman-Ford algorithm iterates *i* steps, it can be interpreted as taking *i* steps from *t* to *v* and the numbers of paths is $O(n^i)$. Each path corresponds to a vector $\boldsymbol{a}(w(r), l(r))$, all the paths with *i* steps from *t* to *v* form the set $U_{vi}$. The hypothesis we propose next is based on Lemma 3.3. Because $l(e)$ are symmetric with each other and the lengths of paths are the sum of lengths of the edges, we hypothesize that all the lengths of paths with *i* steps from *t* to *v* have some symmetries. Similarly, we hypothesize that all the weights of paths with *i* steps from *t* to *v* have some symmetries and even if the symmetries are poor, the increase in complexity of the algorithm they cause is polynomial time, that is, $W_{oR}(U_{vi})$ are mostly in $[1, \lceil O(e \log(n^i - 1)) \cdot p_n(n) \rceil + M_3]$ or $[1, \lceil O(p_n(n) i \log n) \rceil + M_3]$ and the probabilily that $W_{oR}(U_i) > \lceil O(p_e(n) i \ln n) \rceil + M_3$ is less than or equal to $e^{-M_3}$, where $i \in \{1, 2, \ldots, n-1\}$, $p_n(n)$ is a polynomial in *n* and $M_3$ is a large positive integer.

To simplify the analysis process, we change formula (1) to $OPT(i, v) \overset{oR}{\subseteq} \cup_{w \in V}\{OPT(i-1, w) + \boldsymbol{a_{vw}}\}$ and $M(i, v) = OPT(i, v)$ to $M(i, v) = \{OPT(j, v)\}$, where $j \in \{0, 1, , \ldots, i\}$. Paths with different steps are discussed separately and paths with different steps are not compared. In the iteration *i*, only updates the paths with *i* steps and updating the optimization set only counts once when one edge is involved. All *n* points and $O(n)$ edges of each point in *n* iterations, the number of finding optimization set is $O(n^3)$. The probability that the complexity of finding an optimization set is $O(p_n^2(n) n^2 \ln^2 n)$ is greater than or equal to $1 - e^{-M_3}$. The probability that the complexity of the algorithm is $O(p_n^2(n) n^5 \ln^2 n)$ is greater than or equal to $(1 - e^{-M_3})^{n^3}$. We can express the complexity of the algorithm in terms of probability, which is $p(O(p_n^2(n) n^5 \ln^2 n)) \geq (1 - e^{-M_3})^{n^3}$ or denoted by $O_p(p_n^2(n) n^5 \ln^2 n) \geq (1 - e^{-M_3})^{n^3}$. In the expansion of $(1 - e^{-M_3})^{n^3}$, the ratio of the absolute value of the latter term to the former term is less than or equal to $n^3 e^{-M_3}$ which becomes very small as $M_3$ gets larger. So we just focus on the first two terms $1 - n^3 e^{-M_3}$ and the probability still approaches 1 exponentially with the increase of constant term $M_3$. Let $p_e = \max(p_{ei})$, where $i \in \{0, 1, \ldots, n-1\}$ and the complexity of the actual calculation is $O(n^3 p_e^2)$.

Let $\eta = O(n^i)$ and $p^r(\eta, m)$ be the probability of $W_{oR}(U_{vi}) = m$. We show that our hypothesis is resonable by analysing a bad case, in which poor symmetries lead to exponential growth of the probability of larger numbers of $W_{oR}(U_{vi})$, that is, $p^r(\eta, m) = P_1(\eta, m) \cdot \mu^m \cdot f(\eta, \mu)$, where $\mu$ is a constant greater than 1 and $f(\eta, \mu)$ is a function of $\eta$ and $\mu$ to make $\sum_{m=1}^{\eta} p^r(\eta, m) = 1$. We can easily get $f(\eta, \mu) \leq 1$. In proving Theorem 2.4, we change $m \geq$ $\lceil ek\log(n-1) + \frac{e\pi^2}{6}(2^k - 1)\rceil$ to $m \geq \lceil e\mu\log(n-1) + e\mu \rceil$ and others should be changed accordingly. We can get $p^r(\eta, m > M^r) \leq e^{-M_1^r}$, where $M^r = \lceil e\mu\log(\eta - 1) + e\mu \rceil + M_1^r$ and $M_1^r$ is a positive integer. The results of this bad case does not affect the analysis of the algorithm and also accords with the following conjecture.

Based on the above simulation experiments and analysis, we put forward the conjecture that the probability of solving the edge-symmetry and weight-symmetry shortest weight-constrained path problem with *2*-dimensional Bellman-Ford algorithm in polynomial time approaches 1 exponentially with the increase of constant term in algorithm complexity.

## 4 CONCLUSION

In this paper, we define the optimization set and try to extend *1*-dimensional optimization algorithm to multidimensional. We define the sequential optimization numbers and give its recurrence formula, upper bound and some application. We also find the relationship between sequential optimization numbers and the Stirling numbers of first kind. We take advantage of the fact that *1*-dimensional sequential optimization numbers are almost concentrated in $\{O_1(n,\beta)\}$, where $\beta \in [1, \lceil O(\log n) \rceil]$ and apply it to NP-complete problems that currently only have solutions with exponential complexity. We design edge-symmetry and weight-symmetry shortest weight-constrained path problem and multidimensional Bellman-Ford algorithm and conjecture that there is a high probability that the problem can be solved in polynomial time with reasonable analysis.

In physics, there many symmetric cases and there is the phenomenon that in theory one situation, but in fact a particular situation occurs with the probability of almost 1. When errors are allowed in experiment-based physics research, such contradictory phenomena become what we consider to be true, but in fact they are related by probabilistic relationship. If the conjecture is correct, we can conclude something like this. In the real world, there are many symmetric or random situations and similar algorithms can solve similar problems in many application scenarios. This also suggests that in classical complexity analysis, some excluded algorithms can effectively solve problems in special cases.

### ACKNOWLEDGMENTS

All the conclusions of this paper are derived from the process of seeking theoretical proof of *Problem F* of the Huawei Cup 16th China Post-Graduate Mathematical Contest in Modeling in 2019. We thank the organizing committee and the organizations and individuals involved.

# A PROOF OF LEMMA 2.4, INSTANCE 2.1 AND ALL THEOREMS

## A.1 Proof of Theorem 2.1

**Theorem 2.1.** *For all $k, n \in N$,*

$$O_k(n,m) = \begin{cases} (n-1)!^k, & m=1, n>0 \\ (n-1)!^k \sum \prod_{i=1}^{m-1} \dfrac{j_i^k - (j_i-1)^k}{(j_i-1)^k}, & m = \{2,3,\ldots n\} \\ 1, & m=0, n=0 \\ 0, & otherwise \end{cases}$$

*where $j_1, j_2, \cdots, j_{m-1}$ are all m-1-combinations of $\{2,3,\ldots,n\}$.*

**Proof.** In the scenario of Definition 2.2, $g(S)$ denotes the numbers of ways when $S$ is a particular set and $S \subseteq^{OR} U$. $S$ is optimization set when $k = 1$.

First, we discuss the situation when $k = 1$.

For $k = 1$ and $m = 1$, the number of positions of $a_2 \sim a_n$ is $(n-1)!$ as shown in figure 3.a. So $O_1(n,1) = (n-1)!$.

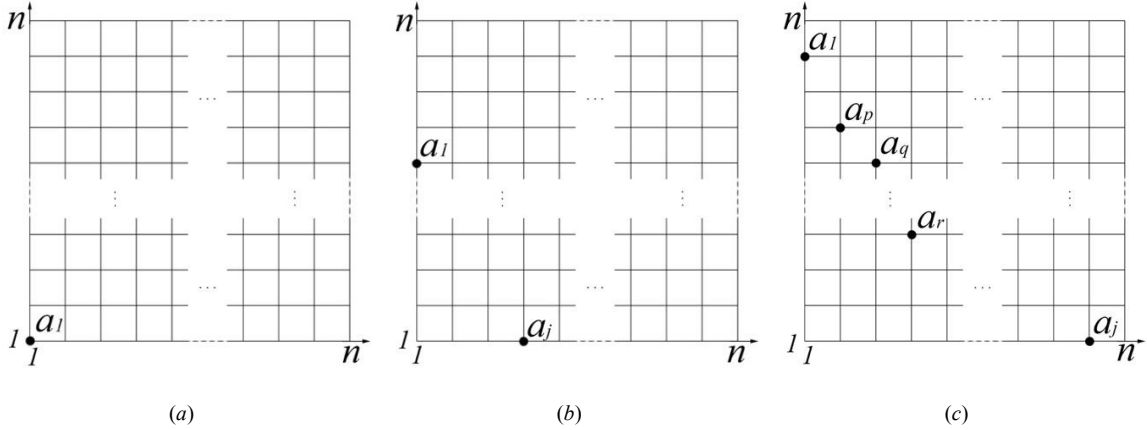

Figure 3: Three ways to optimization set.

For $k=1, m \in \{2,3,\ldots,n\}$, $|U| = n$ and $S_1 = \{a_1, a_p, a_q, \ldots, a_r, a_j\}$, we use mathematical induction to prove that $g(S_1) = \dfrac{(n-1)!}{(p-1)(q-1)\cdots(r-1)(j-1)}$.

Basis step: prove $g(S_2) = \dfrac{(n-1)!}{j-1}$, where $S_2 = \{a_1, a_j\}$. As shown in figure 3.b, $S_2 = \{a_1, a_j\}$, $a_{11} = i$, where $i \in \{2,3,\ldots,n\}$, $a_{j1} = 1$, where $j \in \{2,3,\ldots,n\}$ and $i + j \le n + 2$. First, place row $2 \sim (i-1)$ and the number of positions is $\dfrac{(n-j)!}{(n-i-j+2)!}$. Second, place row $(i+1) \sim n$ and the number of positions is $(n-i)!$. Total number is $\dfrac{(n-i)!(n-j)!}{(n-i-j+2)!}$.

$$g(S_2) = \sum_{i=2}^{n+2-j} \dfrac{(n-i)!(n-j)!}{(n-i-j+2)!}$$

$$= \dfrac{(n-j)!}{j-1} \sum_{i=2}^{n+2-j} \dfrac{(n-i)![(n-i+1)-(n-i-j+2)]}{(n-i-j+2)!}$$

$$= \frac{(n-j)!}{j-1} \sum_{i=2}^{n+2-j} \left[\frac{(n-i+1)!}{(n-i-j+2)!} - \frac{(n-i)!\,(n-i-j+2)}{(n-i-j+2)!}\right]$$

In the brackets, the right side of the minus is equal to 0 when $i = n + 2 - j$ and the right side of the minus when $i = t$ is equal to the left side of the minus when $i = t + 1$, where $t \in \{2,3,\ldots,n+1-j\}$. So,

$$g(S_2) = \frac{(n-j)!}{j-1} \cdot \frac{(n-1)!}{(n-j)!}$$
$$= \frac{(n-1)!}{j-1}$$

Inductive step: assume $g(S_3) = \frac{(n-1)!}{(p-1)(q-1)\cdots(r-1)}$, where $S_3 = \{a_1, a_p, a_q, \ldots, a_r\}$. Then, we show $g(S_1) = \frac{(n-1)!}{(p-1)(q-1)\cdots(r-1)(j-1)}$ where $S_1 = \{a_1, a_p, a_q, \ldots, a_r, a_j\}$, where $1 < p < q < r < j \le n$. In figure 3.c, $a_{r1} = i$, $a_{j1} = 1$, where $i \in \{2,3,\ldots,n-1\}$, $j \in \{3,4,\ldots,n\}$ and $i + j \le n + 2$. First, place row $2 \sim (i-1)$ and the number of positions is $\frac{(n-j)!}{(n-i-j+2)!}$. We remove all the rows and columns where the points in row $1 \sim (i-1)$ are. In the rest of the figure, $|U'| = n - i + 1$, $S'_3 = \{a_1, a_p, a_q, \ldots, a_r\}$ and the number of ways is $\frac{(n-i)!}{(p-1)(q-1)\cdots(r-1)}$. So,

$$g(S_1) = \sum_{i=2}^{n+2-j} \frac{(n-i)!\,(n-j)!}{(p-1)(q-1)\cdots(r-1)(n-i-j+2)!}$$
$$= \frac{(n-1)!}{(p-1)(q-1)\cdots(r-1)(j-1)}$$

So, for $k = 1$, change $S_3$ to $S_1$ by add $a_j$ and we can get $g(S_1) = \frac{1}{j-1} g(S_3)$, where $j \in \{2,3,\ldots,n\}$.

Then, for $k > 1$, we discuss the relationship between $g(S_1)$ and $g(S_3)$. In $a_1 \sim a_n$, let dimension $1$, $\{1, 2, \ldots n\}$ and dimension $w+1$, $\{a_{1w}, a_{2w}, \ldots, a_{nw}\}$ form $k$ groups of $n$ 2-dimensional vectors $(1, a_{1w}), (2, a_{2w}), \ldots (n, a_{nw})$ and $U^w = \{(1, a_{1w}), (2, a_{2w}), \ldots (n, a_{nw})\}$, where $w \in \{1,2,\ldots,k\}$. Let $S^w \overset{OR}{\subseteq} U^w$ and we can get the relationship between $S^w$ and $g(S^w)$ is the same as the relationship between $S$ and $g(S)$ when $k = 1$. In $k+1$-dimensional vector, to change $S_3$ to $S_1$ by add $a_j$, we need to change at least one $S_3^w$ to $S_1^w$ by add $(j, a_{jw})$ and we can get

$$g(S_1) = (C_k^1 \frac{1}{j-1} + C_k^2 \frac{1}{(j-1)^2} + \cdots + C_k^k \frac{1}{(j-1)^k}) g(S_3)$$
$$g(S_1) = \frac{j^k - (j-1)^k}{(j-1)^k} g(S_3)$$

where $j \in \{2,3,\ldots,n\}$. So, for all $k \ge 1$, $O_k(n,1) = (n-1)!^k$ and $O_k(n,m) = (n-1)!^k \sum \prod_{i=1}^{m-1} \frac{j_i^k - (j_i-1)^k}{(j_i-1)^k}$, where $m \in \{2,3,\ldots n\}$ and $j_1, j_2, \cdots, j_{m-1}$ are all $m$-$1$-combinations of $\{2,3,\ldots,n\}$. For $m > n$, we can easily get $O_k(n,m) = 0$.

For all $n = 0$ and $k \ge 0$ we define that

$$O_k(n,m) = \begin{cases} 1, & m = 0 \\ 0, & m > n \end{cases}$$

For all $n > 0$ and $k = 0$ we define that

$$O_k(n,m) = \begin{cases} 1, & m = 1 \\ 0, & otherwise \end{cases}$$

To sum up, for all $k, n \in N$,

$$O_k(n,m) = \begin{cases} (n-1)!^k, & m=1, n>0 \\ (n-1)!^k \sum \prod_{i=1}^{m-1} \frac{j_i^k - (j_i-1)^k}{(j_i-1)^k}, & m \in \{2,3,\ldots n\} \\ 1, & m=0, n=0 \\ 0, & otherwise \end{cases}$$

where $j_1, j_2, \cdots, j_{m-1}$ are all *m-1*-combinations of $\{2,3,\ldots,n\}$.

## A.2 Proof of Theorem 2.2

**Theorem 2.2.** *For all $m, n \in N$ and $k \in N^+$, the recurrence formula of k-dimensional sequential optimization numbers is*

$$O_k(n+1, m+1) = n^k O_k(n, m+1) + [(n+1)^k - n^k] O_k(n, m)$$

*and the boundary condition is*

$$O_k(n,m) = \begin{cases} 0, & m=0, n>0 \text{ or } m>n \\ 1, & m=0, n=0 \end{cases}$$

**Proof.** In the scenario of Definition 2.2, we divide the ways that $O_k(n+1, m+1)$ denote into two parts which are with and without $\boldsymbol{a_{n+1}}$, where $n > 2$ and $m \in \{2,3,\ldots,n-1\}$, First, ways with $\boldsymbol{a_{n+1}}$ and m vectors from $\{\boldsymbol{a_1}, \boldsymbol{a_2}, \ldots, \boldsymbol{a_n}\}$. let $j_m$ be $n+1$ and $j_1, j_2, \cdots, j_{m-1}$ be all *m-1*-combinations of $\{2,3,\ldots,n\}$.

$$n!^k \sum \prod_{i=1}^{m} \frac{j_i^k - (j_i-1)^k}{(j_i-1)^k} = n!^k \frac{(n+1)^k - n^k}{n^k} \sum \prod_{i=1}^{m-1} \frac{j_i^k - (j_i-1)^k}{(j_i-1)^k}$$

$$= n!^k \frac{(n+1)^k - n^k}{n^k} \sum \prod_{i=1}^{m-1} \frac{j_i^k - (j_i-1)^k}{(j_i-1)^k}$$

$$= [(n+1)^k - n^k](n-1)!^k \sum \prod_{i=1}^{m-1} \frac{j_i^k - (j_i-1)^k}{(j_i-1)^k}$$

$$= [(n+1)^k - n^k] O_k(n, m)$$

Then, ways without $\boldsymbol{a_{n+1}}$ and m vectors from $\{\boldsymbol{a_1}, \boldsymbol{a_2}, \ldots, \boldsymbol{a_n}\}$, $j_1, j_2, \cdots, j_m$ are all *m*-combinations of $\{2,3,\ldots,n\}$.

$$n!^k \sum \prod_{i=1}^{m} \frac{j_i^k - (j_i-1)^k}{(j_i-1)^k} = n^k (n-1)!^k \sum \prod_{i=1}^{m} \frac{j_i^k - (j_i-1)^k}{(j_i-1)^k}$$

$$= n^k O_k(n, m+1)$$

To sum up,

$$O_k(n+1, m+1) = n^k O_k(n, m+1) + [(n+1)^k - n^k] O_k(n, m)$$

We can get the boundary condition according to Theorem 2.1.

$$O_k(n,m) = \begin{cases} 0, & m=0, n>0 \text{ or } m>n \\ 1, & m=0, n=0 \end{cases}$$

We can prove that the recurrence formula works for all m, n ∈ N and k ∈ N⁺. We also notice that it works for all $m, n, k \in N$, if we define $0^0 = 0$.

## A.3 Proof of Instance 2.1

**Instance 2.1.** *There are n boards and the sequence of them are fixed. Each board is painted one color and divided into k smaller boards. All the ith smaller boards of each board form a group and their height are $1, 2, \ldots, n$ respectively, where $i \in \{1, 2, \ldots, k\}$. Following the direction of the arrow, the number of ways that m colors can be seen is $O_k(n, m)$.*

**Proof.** Label as the $i$th board follow direction of the arrow. Label the height of the smaller boards in same color from left to right as $h_{i1}, h_{i2}, \ldots h_{ik}$. Let $\mathbf{R} = (>, >, \ldots, >)$ be $k$-dimensional relation vector, $\mathbf{h_i} = (i, h_{i1}, h_{i2}, \ldots h_{ik})$ and $a_{ij} = n + 1 - h_{ij}$, where $i \in \{1, 2, \ldots, n\}$ and $j \in \{1, 2, \ldots, k\}$. When $k = 1$, as shown in Figure 4, we can get same optimization set in Figure 3.c follow direction of the line of sight. When $k \geq 1$, a certain color is seen equals that color is seen in at least one group. So, we can conclude that the number of ways that $m$ colors can be seen is $O_k(n, m)$.

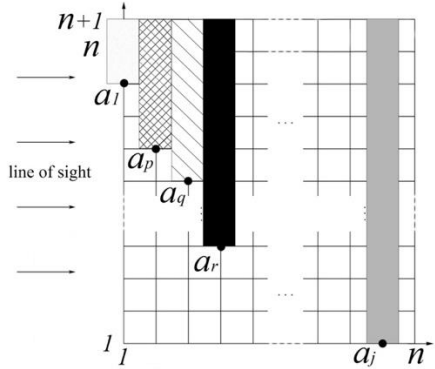

Figure 4: Relationship between smaller color boards group and optimization set.

## A.4 Proof of Theorem 2.3

**Theorem 2.3.** *For all $k \geq 1, n \geq 2$ and $m \geq 1$, we define*

$$O_{kmax}(n, m) = \frac{(n-1)!^k}{(m-1)!} \{\sum_{i=1}^{k} [C_k^i \cdot H_i(n-1)]\}^{m-1}$$

*where $H_i(n-1) = \sum_{j=1}^{n-1} \frac{1}{j^i}$. We can get $O_{kmax}(n, m) \geq O_k(n, m)$.*

**Proof.** We prove it using Theorem 2.2. First, we prove the boundary condition.

$$O_{kmax}(n, m) \begin{cases} = (n-1)!^k, & m = 1 \\ \geq 0, & m > n \end{cases}$$

So, the boundary condition satisfy $O_{kmax}(n, m) \geq O_k(n, m)$.

Then, we prove the inequality for $2 \leq m \leq n$ by mathematical induction.

Basis step: For $n = 2$,

$$O_{kmax}(2, 2) = 2^k - 1 = O_k(2, 2)$$

So, for $n = 2$, $O_{kmax}(n, m) \geq O_k(n, m)$.

Inductive step: For $n \geq 2$ and $m \geq 2$, we assume $O_{kmax}(n,m) \geq O_k(n,m)$. Then, we prove $O_{kmax}(n+1,m) \geq O_k(n+1,m)$.

$$O_{kmax}(n+1,m) = \frac{n!^k}{(m-1)!}\{\sum_{i=1}^{k}[C_k^i \cdot H_i(n)]\}^{m-1}$$

$$= \frac{n!^k}{(m-1)!}\{\sum_{i=1}^{k}[C_k^i \cdot H_i(n-1)] + \sum_{i=1}^{k}[C_k^i \cdot \frac{1}{n^i}]\}^{m-1}$$

$$\geq \frac{n!^k}{(m-1)!}C_{m-1}^0\{\sum_{i=1}^{k}[C_k^i \cdot H_i(n-1)]\}^{m-1} + \frac{n!^k}{(m-1)!}C_{m-1}^1 \sum_{i=1}^{k}[C_k^i \cdot \frac{1}{n^i}]\{\sum_{i=1}^{k}[C_k^i \cdot H_i(n-1)]\}^{m-2}$$

$$= n^k \frac{(n-1)!^k}{(m-1)!}\{\sum_{i=1}^{k}[C_k^i \cdot H_i(n-1)]\}^{m-1} + \frac{n!^k}{(m-2)!}[(\frac{1}{n}+1)^k - 1]\{\sum_{i=1}^{k}[C_k^i \cdot H_i(n-1)]\}^{m-2}$$

$$= n^k \frac{(n-1)!^k}{(m-1)!}\{\sum_{i=1}^{k}[C_k^i \cdot H_i(n-1)]\}^{m-1} + [(n+1)^k - n^k]\frac{(n-1)!^k}{(m-2)!}\{\sum_{i=1}^{k}[C_k^i \cdot H_i(n-1)]\}^{m-2}$$

$$= n^k O_{kmax}(n,m) + [(n+1)^k - n^k]O_{kmax}(n,m-1)$$

$$\geq n^k O_k(n,m) + [(n+1)^k - n^k]O_k(n,m-1)$$

$$= O_k(n+1,m)$$

To sum up, For all $k \geq 1, n \geq 2$ and $m \geq 1$, $O_{kmax}(n,m) \geq O_k(n,m)$.

### A.5 Proof of Theorem 2.4

**Theorem 2.4.** *Let $P_k(n,m) = \frac{O_k(n,m)}{n!^k}$ be the probability of $O_k(n,m)$ and $M = \lceil ek \log(n-1) + \frac{e\pi^2}{6}(2^k-1) \rceil + M_1$, where $M_1$ is a positive integer. For all $n \geq 2$ and $k \geq 1$, we can get*

$$P_k(n, m > M) \leq e^{-M_1}$$

**Proof.** Let $P_{kmax}(n,m) = \frac{O_{kmax}(n,m)}{n!^k}$ be the probability of the upper bound of sequential optimization numbers.

$$P_{kmax}(n,m) = \frac{O_{kmax}(n,m)}{n!^k}$$

$$= \frac{1}{n^k(m-1)!}\{\sum_{i=1}^{k}[C_k^i \cdot H_i(n-1)]\}^{m-1}$$

and,

$$\frac{P_{kmax}(n,m+1)}{P_{kmax}(n,m)} = \frac{\sum_{i=1}^{k}[C_k^i \cdot H_i(n-1)]}{m}$$

We know that $H_1(n-1) \leq \log(n-1) + 1$, $m! > \sqrt{2\pi m}(\frac{m}{e})^m$, and for $i \geq 2$, $H_i(n-1) \leq \frac{\pi^2}{6}$. So,

$$\frac{P_{kmax}(n, m+1)}{P_{kmax}(n, m)} \leq \frac{k \log(n-1) + k + \sum_{i=2}^{k}[C_k^i \cdot \frac{\pi^2}{6}]}{m}$$

$$\leq \frac{k \log(n-1) + k + \frac{\pi^2}{6}(2^k - k - 1)}{m}$$

$$\leq \frac{k \log(n-1) + \frac{\pi^2}{6}(2^k - 1)}{m}$$

When $m \geq \lceil ek \log(n-1) + \frac{e\pi^2}{6}(2^k - 1) \rceil$,

$$\frac{P_{kmax}(n, m+1)}{P_{kmax}(n, m)} \leq \frac{1}{e}$$

and,

$$P_{kmax}(n, m+1) = \frac{1}{n^k m!} \{\sum_{i=1}^{k}[C_k^i \cdot H_i(n-1)]\}^m$$

$$\leq \frac{[k \log(n-1) + \frac{\pi^2}{6}(2^k - 1)]^m}{n^k \sqrt{2\pi m}(\frac{m}{e})^m}$$

$$= \frac{1}{n^k \sqrt{2\pi m}}[\frac{ek \log(n-1) + \frac{e\pi^2}{6}(2^k - 1)}{m}]^m$$

$$\leq e^{-1}$$

When $M = \lceil ek \log(n-1) + \frac{e\pi^2}{6}(2^k - 1) \rceil + M_1$, where $M_1$ is a positive integer.

$$P_{kmax}(n, M) \leq e^{-M_1}$$

So,

$$P_k(n, m > M) \leq P_{kmax}(n, m > M)$$

$$= \sum_{i=M+1}^{n} P_{kmax}(n, i)$$

$$\leq e^{-M_1}$$

### A.6 Proof of Lemma 2.4

**Lemma 2.4.** *In particular, For $k = 1$, $M = \lceil e \log(n-1) + e \rceil + M_1$, $P_1(n, m > M) \leq e^{-M_1}$. $O_1(n, m)$, or $s_u(n, m)$, are mostly concentrated in $\{O_1(n, \alpha)\}$, where $\alpha \in [1, \lceil e \log(n-1) + e \rceil + M_1]$. Let $M_2 = M_1 + 3$ and $O_1(n, m)$ are mostly concentrated in $\{O_1(n, \alpha)\}$, where $\alpha \in [1, \lceil e \log(n-1) \rceil + M_2]$ and $M_2$ is a large positive integer.*

**Proof.** In proving Theorem 2.4, we can prove Lemma 2.4 by replace

$$\frac{\sum_{i=1}^{k}[C_k^i \cdot H_i(n-1)]}{m} \leq \frac{k\log(n-1) + \sum_{i=1}^{k}[C_k^i \cdot \frac{\pi^2}{6}]}{m}$$

with

$$\frac{\sum_{i=1}^{1}[C_1^i \cdot H_i(n-1)]}{m} \leq \frac{\log(n-1) + 1}{m}$$

and $m \geq \lceil ek\log(n-1) + \frac{e\pi^2}{6}(2^k - 1) \rceil$ with $m \geq \lceil e\log(n-1) + e \rceil$.

### A.7 Proof of Theorem 2.5

**Theorem 2.5.** *For all $k \geq 1, n \geq 2$,*

$$\frac{\sum_{i=1}^{n} O_{kmax}(n,i)}{\sum_{i=1}^{n} O_k(n,i)} \leq (\frac{n-1}{n})^k e^{\frac{\pi^2}{6}(2^k - 1)}$$

*In particular, for $k = 1, n \geq 2$,*

$$\frac{\sum_{i=1}^{n} O_{1max}(n,i)}{\sum_{i=1}^{n} O_1(n,i)} \leq \frac{(n-1)}{n} e^{H_1(n-1) - \log(n-1)} \leq e$$

**Proof.** Let

$$\lambda = \sum_{i=1}^{k} [C_k^i \cdot H_i(n-1)]$$

$$\leq k\log(n-1) + k + \frac{\pi^2}{6}(2^k - k - 1)$$

then,

$$\frac{\sum_{i=1}^{n} O_{kmax}(n,i)}{\sum_{i=1}^{n} O_k(n,i)} = \frac{1}{n!^k} \sum_{i=1}^{n} \frac{(n-1)!^k}{(i-1)!} \lambda^{i-1}$$

$$= \frac{1}{n^k} \sum_{i=1}^{n} \frac{\lambda^{i-1}}{(i-1)!}$$

$$\leq \frac{e^\lambda}{n^k} \quad \text{(Taylor theorem)}$$

$$\leq \frac{e^{k\log(n-1) + k + \frac{\pi^2}{6}(2^k - k - 1)}}{n^k}$$

$$\leq (\frac{n-1}{n})^k e^{\frac{\pi^2}{6}(2^k - 1)}$$

In particular, for $k = 1$,

$$\frac{\sum_{i=1}^{n} O_{1max}(n,i)}{\sum_{i=1}^{n} O_1(n,i)} \leq \frac{(n-1)}{n} e^{H_1(n-1)-\log(n-1)} \leq e$$

We can get a smaller M and similar properties of Theorem 2.4 and Lemma 2.4 by using the conclusions of Theorem 2.5 and selecting a reasonable ratio of $\frac{P_{kmax}(n,m+1)}{P_{kmax}(n,m)}$.

## B  SOME RESULTS OF SEQUENCES AND EXPERIMENT

For $n \leq 6$, according to recurrence formula, 2-dimensional sequential optimization numbers are shown in Table 1 and 3-dimensional sequential optimization numbers are shown in Table 2.

Table 1: 2-dimensional sequential optimization numbers

| 1 | 0 | 0 | 0 | 0 | 0 | 0 |
|---|---|---|---|---|---|---|
| 0 | 1 | 0 | 0 | 0 | 0 | 0 |
| 0 | 1 | 3 | 0 | 0 | 0 | 0 |
| 0 | 4 | 17 | 15 | 0 | 0 | 0 |
| 0 | 36 | 181 | 254 | 105 | 0 | 0 |
| 0 | 576 | 3220 | 5693 | 3966 | 945 | 0 |
| 0 | 14400 | 86836 | 177745 | 161773 | 67251 | 10395 |

Table 2: 3-dimensional sequential optimization numbers

| 1 | 0 | 0 | 0 | 0 | 0 | 0 |
|---|---|---|---|---|---|---|
| 0 | 1 | 0 | 0 | 0 | 0 | 0 |
| 0 | 1 | 7 | 0 | 0 | 0 | 0 |
| 0 | 8 | 75 | 133 | 0 | 0 | 0 |
| 0 | 216 | 2321 | 6366 | 4921 | 0 | 0 |
| 0 | 13824 | 161720 | 549005 | 703270 | 300181 | 0 |
| 0 | 1728000 | 21472984 | 83342145 | 137868205 | 101520195 | 27316471 |

Increase the number of weights in edge-symmetry and weight-symmetry shortest weight-constrained path problem and the edges and all weights are still symmetric with each other separately. Multidimensional Bellman-Ford algorithm can be obtained by extending the 2-dimensional Bellman-Ford algorithm. Take the graph in Figure 2.a and the relationship between $i$ and $p_{ei}$ of a 3-dimensional simulation experiment is shown in Figure 5.a and a 4-dimensional simulation experiment is shown in Figure 5.b. The relationship between $i$ and $p_{ei}$ is not like exponential and the growth of $p_{ei}$ is still acceptable.

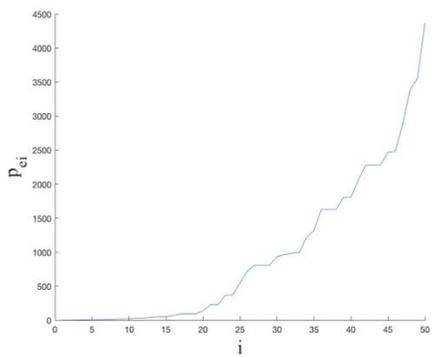 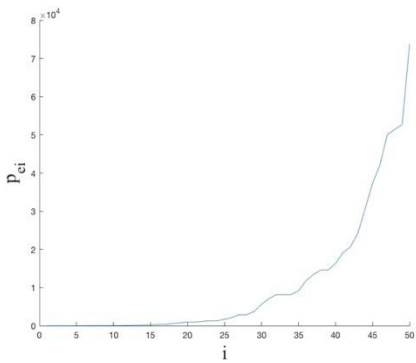

(*a*)  (*b*)

Figure 5: Simulation experiments of *3* and *4*-dimensional Bellman-Ford algorithm.